\documentclass[amsmath,floatfix,twocolumn,superscriptaddress]{revtex4-1}
\usepackage{amssymb,amsmath}
\usepackage{graphicx,array}
\usepackage{dcolumn}
\usepackage{psfrag}
\usepackage{bm}
\usepackage{color}

\begin{document}
\title{Prediction of large barocaloric effects in thermoelectric superionic materials}

\author{Jie Min}
\affiliation{School of Materials Science and Engineering, UNSW Sydney, Sydney, NSW 2052, Australia} 

\author{Arun K. Sagotra}
\affiliation{School of Materials Science and Engineering, UNSW Sydney, Sydney, NSW 2052, Australia} 

\author{Claudio Cazorla}
\affiliation{School of Materials Science and Engineering, UNSW Sydney, Sydney, NSW 2052, Australia}

\begin{abstract}
We predict the existence of large barocaloric effects above room temperature in the thermoelectric 
fast-ion conductor Cu$_{2}$Se by using classical molecular dynamics simulations and first-principles 
computational methods. A hydrostatic pressure of $1$~GPa induces large isothermal entropy changes of 
$|\Delta S| \sim 15$--$45$~J~kg$^{-1}$~K$^{-1}$ and adiabatic temperature shifts of $|\Delta T| \sim 10$~K 
in the temperature interval $400 \le T \le 700$~K. Structural phase transitions are absent in the 
analysed thermodynamic range. The causes of such large barocaloric effects are significant $P$--induced
variations on the ionic conductivity of Cu$_{2}$Se and the inherently high anharmonicity of the material. 
Uniaxial stresses of the same magnitude, either compressive or tensile, produce comparatively much smaller 
caloric effects, namely, $|\Delta S| \sim 1$~J~kg$^{-1}$~K$^{-1}$ and $|\Delta T| \sim 0.1$~K, due to 
practically null influence on the ionic diffusivity of Cu$_{2}$Se. Our simulation work shows that 
thermoelectric compounds presenting high ionic disorder, like copper and silver-based chalcogenides, 
may render large mechanocaloric effects and thus are promising materials for engineering solid-state 
cooling applications that do not require the application of electric fields. 
\end{abstract}
\maketitle

\section{Introduction}
\label{sec:intro}
Conventional cooling technologies are based on compression cycles of greenhouse gases (e.g., 
hydrofluorocarbons), which pose serious threats to the environment. One kilogram of a typical 
refrigerant gas is, in terms of greenhouse impact, equivalent to two tonnes of carbon dioxide, 
which equals the total emissions produced by a car running uninterruptedly during six months 
\cite{cazorla19a}. Current cooling technologies, in addition, present two other important 
limitations, namely, the energy efficiency of the refrigeration cycles are relatively low 
($< 60$\%) \cite{moya14}, and they cannot be scaled down to small sizes (e.g., microchip 
dimensions).

Solid-state cooling is an emergent refrigeration technology that exploits thermal effects in 
materials, and which could solve most of the problems associated with traditional vapour-compression
refrigeration. For instance, solid-state cooling systems do not represent any environmental threat 
in terms of greenhouse gas emissions and in principle can be reduced in size to fit within portable 
devices. The absence of moving parts and silent operation represent additional advantages 
over traditional refrigeration technologies. Two broad families of materials that are employed 
in solid-state cooling applications are caloric \cite{moya14} and thermoelectric \cite{zhao14} 
compounds.  

Caloric materials react thermally to external coercive fields like electric and magnetic bias 
\cite{moya14,cazorla18b,krenke05,defay13} and mechanical stresses \cite{manosa17,cazorla17b,tusek15}. 
Caloric effects result from field-induced transformations that involve large changes in entropy 
($\sim 10$--$100$~Jkg$^{-1}$K$^{-1}$). One typical example of caloric materials are shape-memory
alloys (e.g., near equiatomic Ni-Ti alloys), which exhibit superb elastic properties and a martensite 
to austenite phase transition with a large latent heat that can be triggered by external fields 
\cite{manosa17,tusek15}. Solid-state cooling energy efficiencies of $\sim 75$\% have been demonstrated 
for some caloric materials and further improvements appear to be within reach \cite{moya14,zimm98}. 

Meanwhile, thermoelectric materials create an electric potential when subjected to a temperature gradient 
and \emph{vice versa}, that is, they generate a temperature gradient when subjected to an electric bias. 
Thermoelectric refrigerators exploit the latter effect, known as the Peltier effect, and thus cooling 
is achieved via the application of electric fields \cite{zhao14,snyder02}. The efficiency of thermoelectric 
materials is measured by a dimensionless parameter called ``thermoelectric figure of merit'', which 
typically adopts values of $\sim 1$ for good specimens \cite{zevalkink18}. Unfortunately, the energy 
efficiency of current thermoelectric refrigerators are relatively low as compared to that of conventional 
vapour compression systems \cite{tassou10}. Actually, much higher thermoelectric figures of merit than 
$\sim 1$ are necessary for thermoelectric coolers to become commercially viable \cite{tassou10}. 

Copper selenide, Cu$_{2}$Se, is an inorganic compound for which recently huge thermoelectric figures 
of merit of $\sim 2$ have been reported experimentally at high temperatures ($\sim 1,000$~K) \cite{liu12}. 
Above room temperature, the copper ions in Cu$_{2}$Se become highly mobile and the system enters a 
``superionic'' state \cite{hull04,danilkin12} that is characterised by a very low lattice thermal 
conductivity ($\sim 1$~Wm$^{-1}$K$^{-1}$) \cite{liu12,kim15}. The resulting ``liquid-like'' behaviour 
of the Cu ions appears to be the main cause for the huge thermodynamic figure of merit observed in 
this material. Likewise, large thermoelectric figures of merit have been reported for analogous superionic 
Cu- and Ag-based chalcogenides like Cu$_{2}$S, Cu$_{2}$Te, Ag$_{2}$Se, Ag$_{2}$S, Ag$_{2}$Te, and 
Cu$_{2-x}$Ag$_{x}$$X$ ($X$ = S, Se and Te) alloys \cite{ballikaya13,brown13,han16}. Nevertheless, 
most of such thermoelectric fast-ion conductors exhibit low thermodynamic stability when subjected 
to strong electric fields owing to undesired electromigration \cite{bailey17,dennler14}. Consequently, 
these materials are not suitable for engineering practical solid-state cooling applications based 
on the Peltier effect due to limiting degradation issues \cite{dennler14}. 

In this study, we present theoretical evidence showing that thermoelectric superionic materials 
typified by Cu$_{2}$Se may exhibit large caloric effects when subjected to hydrostatic pressure (that 
is, barocaloric effects). In particular, an isotropic compression of $\sim 1$~GPa in the temperature 
interval $400 \le T \le 700$~K induces a significant decrease in the ionic diffusivity of Cu$_{2}$Se, 
which translates into a large decrease in lattice entropy and overall enhancement in the thermodynamic 
stability of the system. Anharmonicity, which is inherently high in superionic materials 
\cite{cazorla18a,cazorla19,cazorla14}, also plays an important role on the estimated large entropy 
variations. The adiabatic temperature shifts accompanying the barocaloric effects amount to $\sim 10$~K, 
which are reasonably large as compared to those reported for other known barocaloric materials. 
Caloric effects produced by moderate uniaxial stresses (that is, elastocaloric effects), on the contrary, 
are quite small ($\sim 0.1$~K) due to practically negligible influence on the ionic diffusivity of Cu$_{2}$Se. 
Our study suggests that thermoelectric materials presenting high ionic disorder could be used for 
engineering solid-state cooling applications that do not require the application of electric fields, 
thus getting rid of potential thermodynamic instability issues.

\begin{figure*}[t]
\centerline{
\includegraphics[width=1.00\linewidth]{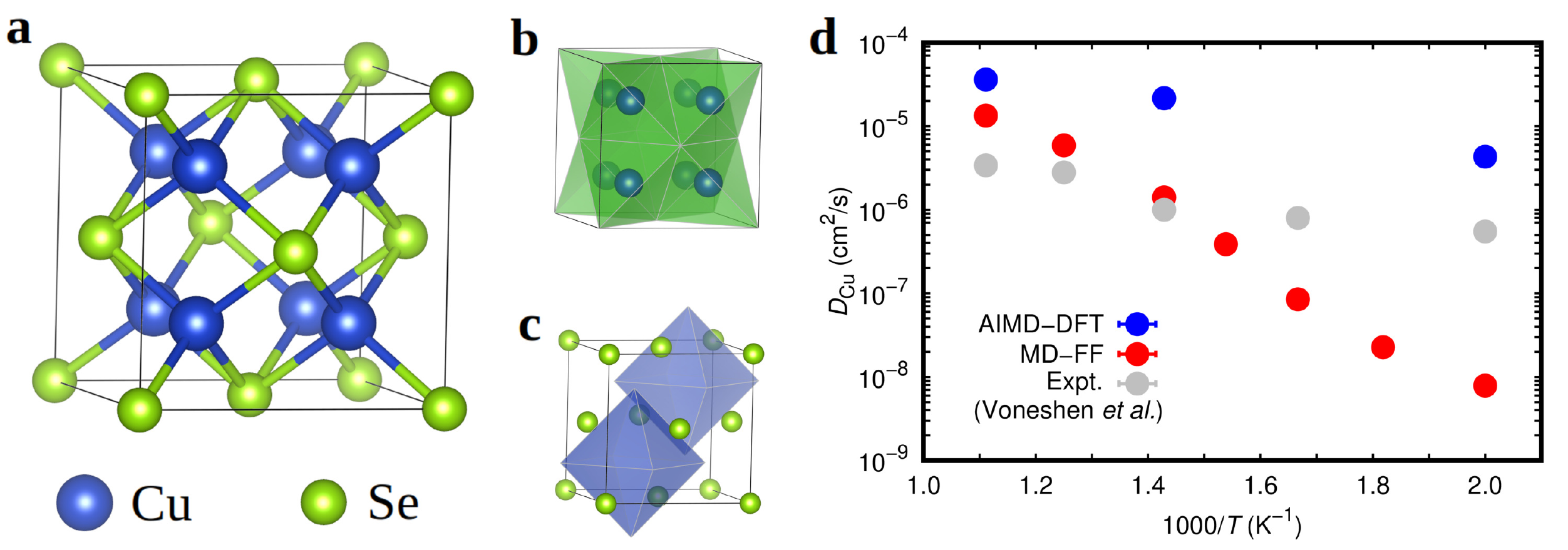}}
\caption{General description of Cu$_{2}$Se. {\bf a} Ball-stick representation of the high-$T$ cubic structure 
	known as fluorite (space group $Fm\overline{3}m$); for clarity purposes, the Cu ions are represented 
	orderly in a simple cubic lattice {\bf b} Tetrahedra formed by Se ions on the vertices;
	the center of the tetrahedra are described by the pseudocubic Wyckoff position $8c$ $\left( \frac{1}{4}, 
	\frac{1}{4}, \frac{1}{4} \right)$. {\bf c} Octahedra formed by Cu ions on the vertices; the center of the 
	octahedra are described by the pseudocubic Wyckoff position $4b$ $\left( \frac{1}{2}, \frac{1}{2}, 
	\frac{1}{2} \right)$. {\bf d} Ionic diffusion coefficients estimated at different temperatures from 
	neutron spectroscopy experiments \cite{voneshen17}, first-principles calculations based on density 
	functional theory (AIMD-DFT), and molecular dynamics simulations performed with the Morse potential 
	reported in \cite{namsani17} considering null ionic charges (MD-FF).}
\label{fig1}
\end{figure*}

\section{Computational methods}
\label{sec:methods}
Molecular dynamics (MD) $( N, P, T )$ simulations were performed with the LAMMPS code \cite{lammps}. 
The pressure and temperature in the system were kept fluctuating around a set-point value by using 
thermostatting and barostatting techniques in which some dynamic variables are coupled to the particle 
velocities and simulation box dimensions. The interactions between atoms were modeled with the Morse 
potential reported by Namsani \emph{et al.} in work \cite{namsani17}. This potential reproduces within 
a few percent the structural and elastic properties of Cu$_{2}$Se at high temperatures as reported 
from experiments and first-principles calculations \cite{namsani17}. The Coulombic interactions between
ions, however, were neglected in this study in order to reproduce correctly the superionic behavior of 
Cu$_{2}$Se at high temperatures (Sec.\ref{sec:results}). 

We employed large simulation boxes, typically containing $14,000$ atoms, and applied periodic boundary 
conditions along the three Cartesian directions. Off-stoichiometric Cu$_{2}$Se configurations were generated 
by removing a specific number of cations and anions randomly from the simulation cell (in order to fulfil 
the condition of charge neutrality). Newton's equations of motion were integrated using the customary Verlet's 
algorithm with a time-step length of $10^{-3}$~ps. The typical duration of a MD run was of $1$~ns. A 
particle-particle particle-mesh $k$-space solver was used to compute long-range interactions beyond a 
cut-off distance of $12$~\AA~ at each time step. MD simulations were performed in the thermodynamic 
intervals $400 \le T \le 800$~K and $0 \le P \le 1$~GPa by considering thermodynamic variable increments 
of $\delta T = 25$~K and $\delta P = 0.1$~GPa.

\emph{Ab initio} molecular dynamics (AIMD) simulations based on density functional theory (DFT) were performed 
to analyse the ionic transport properties of Cu$_{2}$Se and test the reliability of the employed interaction 
potential. These simulatons were performed in the canonical ensemble $( N, V, T )$ with the VASP code \cite{vasp} 
by following the generalized gradient approximation to the exchange-correlation energy due to Perdew \emph{et al.} 
\cite{pbe96}. The projector augmented-wave method was used to represent the ionic cores \cite{bloch94}, and the 
electronic states Cu $4s$-$3d$ and Se $4s$-$4p$ were considered as valence. Wave functions were represented 
in a plane-wave basis truncated at $650$~eV. The temperature in the AIMD simulations was kept fluctuating around 
a set-point value by using Nose-Hoover thermostats. Simulation boxes containing $288$ atoms were used in all the 
AIMD simulations and periodic boundary conditions were applied along the three Cartesian directions. Newton's equations 
of motion were integrated using the customary Verlet's algorithm and a time-step length of $10^{-3}$~ps. $\Gamma$-point 
sampling for integration within the first Brillouin zone was employed in all the AIMD simulations. The total duration 
of each AIMD run was $\sim 200$~ps.

Ionic diffusion coefficients, $D_{\rm Cu}$ and $D_{\rm Se}$, were calculated with the formula \cite{cazorla19} : 
\begin{equation}
D_{i} = \lim_{t \to \infty} \frac{ \langle | R_{i} (t + t_{0}) - R_{i} (t_{0}) |^{2} \rangle }{6 t}~, 
\end{equation}
where $R_{i}(t)$ is the position of the migrating ion labelled as $i$ at time $t$, $t_{0}$ an arbitrary
time origin, and $\langle \cdots \rangle$ denotes average over time and particles. The mean squared 
displacement of each ionic species is defined as $\langle \Delta R_{i}^{2}(t) \rangle \equiv \langle 
| R_{i} (t + t_{0}) - R_{i} (t_{0}) |^{2} \rangle$. 
 
Isothermal entropy changes associated to the barocaloric effect were estimated as \cite{moya14}:
\begin{eqnarray}
\Delta S (P_{\rm max}, T) =  - \int_{0}^{P_{\rm max}} \left(\frac{\partial V}{\partial T}\right)_{P} dP~,
\label{eq:baro}
\end{eqnarray}
where $P_{\rm max}$ represents the maximum applied hydrostatic pressure, and $V$ the volume of the system. 
In the case of elastocaloric effects, the same quantity is calculated as \cite{moya14}:
\begin{eqnarray}
\Delta S (\sigma_{\rm max}, T) = V_{0} \cdot \int_{0}^{\sigma_{\rm max}} \left(\frac{\partial \epsilon}{\partial T}\right)_{\sigma} d\sigma~,
\label{eq:elasto}
\end{eqnarray}
where $\sigma_{\rm max}$ represents the maximum uniaxial stress applied along an arbitrary Cartesian direction 
(denoted here as $z$), $\epsilon$ the strain deformation that the system undergoes along the same direction (i.e., 
$\epsilon (\sigma, T) \equiv \frac{L_{z}(\sigma, T) - L_{z}(0,T)}{L_{z}(0,T)}$ where $L_{z}$ corresponds to the 
length of the simulation box along the Cartesian $z$ direction), and $V_{0}$ the equilibrium volume of the system. 
Finally, the resulting adiabatic temperature shifts were estimated with the formula:
\begin{equation}
\Delta T (\sigma_{f}, T) = - \int_{0}^{\sigma_{f}} \frac{T}{C_{\sigma} (T)} \cdot dS~, 
\label{eq:deltat}
\end{equation}
where $C_{\sigma} (T) = \left( \frac{dU}{dT} \right)_{\sigma}$ is the heat capacity of the crystal, which 
has been calculated at each specific set of stress--temperature conditions.

\section{Results}
\label{sec:results}
Copper selenide exhibits two crystalline phases, a low-temperature phase ($\alpha$) that is stable up to 
$\sim 400$~K and a high-temperature phase ($\beta$) that is superionic \cite{clark70}. The exact structure 
of the $\alpha$ phase still remains under debate \cite{liu13,gulay11} however recent works have proposed 
that is monoclinic with space group $P2_{1}/c$ \cite{nguyen13,chi14}. The $\beta$ phase exhibits the well-known 
cubic fluorite structure found in many binary fast-ion conductors (e.g., CaF$_{2}$ and UO$_{2}$) \cite{cazorla14,cazorla16}, 
in which the Se ions are arranged according to a face-centered cubic lattice (space group $Fm\overline{3}m$) 
(Fig.\ref{fig1}a). Copper ions in the $\beta$ phase diffuse throughout the crystal by hopping between 
tetrahedral sites and off-centered octahedral interstitial positions \cite{danilkin12,cazorla18c,cazorla17} 
(Fig.\ref{fig1}b--c). The critical temperature of the $\alpha$~$\to$~$\beta$ phase transition can be modified 
in practice with alloying and hydrostatic pressure as well \cite{clark70}. 

Figure~\ref{fig1}d shows the Cu diffusion coefficients, $D_{\rm Cu}$, measured recently by Voneshen \emph{et al.} 
at different temperatures by using neutron spectroscopy techniques \cite{voneshen17}. $D_{\rm Cu}$ values of 
$\sim 10^{-7}$--$10^{-6}$~cm$^{2}$s$^{-1}$ have been reported within the temperature interval $500 \le T \le 900$~K.
Our theoretical first-principles results obtained from \emph{ab initio} molecular dynamics (AIMD) simulations systematically 
overestimate those experimental $D_{\rm Cu}$ values by roughly one order of magnitude. For instance, at $T = 500$~K~($900$~K) 
Voneshen \emph{et al.} report $5.5 \cdot 10^{-7}$~($3.4 \cdot 10^{-6}$)~cm$^{2}$s$^{-1}$ whereas we obtain $3.8 \cdot 
10^{-6}$~($4.2 \cdot 10^{-5}$)~cm$^{2}$s$^{-1}$. It should be mentioned, however, that in a previous experimental quasi-elastic 
neutron scattering study by Danilkin \emph{et al.} much larger $D_{\rm Cu}$ values than found by Voneshen \emph{et al.} 
were reported at temperatures close to the $\alpha$~$\to$~$\beta$ transition \cite{danilkin12}. Specifically, a Cu diffusion 
coefficient of $6.1 \cdot 10^{-5}$~cm$^{2}$s$^{-1}$ was measured at $T = 430$~K \cite{danilkin12}, which is significantly
higher than our estimations based on AIMD simulations. Hence, there seems to be a lack of quantitative agreement between 
the sets of experimental $D_{\rm Cu}$ data reported to date for $\beta$--Cu$_{2}$Se, and thus probably additional experiments 
are necessary to resolve such discrepancies.  

We performed molecular dynamics (MD) $( N, P, T )$ simulations with the Morse potential proposed by Namsani \emph{et 
al.} for $\beta$--Cu$_{2}$Se \cite{namsani17}, considering both stoichiometric and off-stoichiometric systems. 
Unexpectedly, we found that for all the analysed compositions the reproduced ionic diffusivities were practically null 
($D_{\rm Cu} < 10^{-9}$~cm$^{2}$s$^{-1}$) at temperatures as high as $1,000$~K. 

Incidentally, we realized that by reducing the value of the ionic charges the diffusion of the copper ions increased. 
Actually, the best agreement between our MD results (considering a realistic off-stoichiometry of $2$\%) and the sets 
of experimental and AIMD $D_{\rm Cu}$ data was obtained for neutrally charged particles (Fig.\ref{fig1}d). In that 
particular case the Cu diffusion coefficients estimated with MD at $T = 500$~K and $900$~K, for instance, are $7.9 \cdot 
10^{-9}$ and $1.3 \cdot 10^{-5}$~cm$^{2}$s$^{-1}$, respectively. In view of these outcomes, and of the importance of 
accounting for ionic disorder in the simulation of superionic Cu$_{2}$Se, we decided to perform the subsequent analysis 
of mechanocaloric effects by adopting null ionic charges. As it is explained below, we do not expect that such a modification 
of the original interatomic potential will introduce significant bias on the determination of realistic mechanocaloric 
effects in $\beta$--Cu$_{2}$Se.

\begin{figure*}[t]
\centerline{
\includegraphics[width=1.00\linewidth]{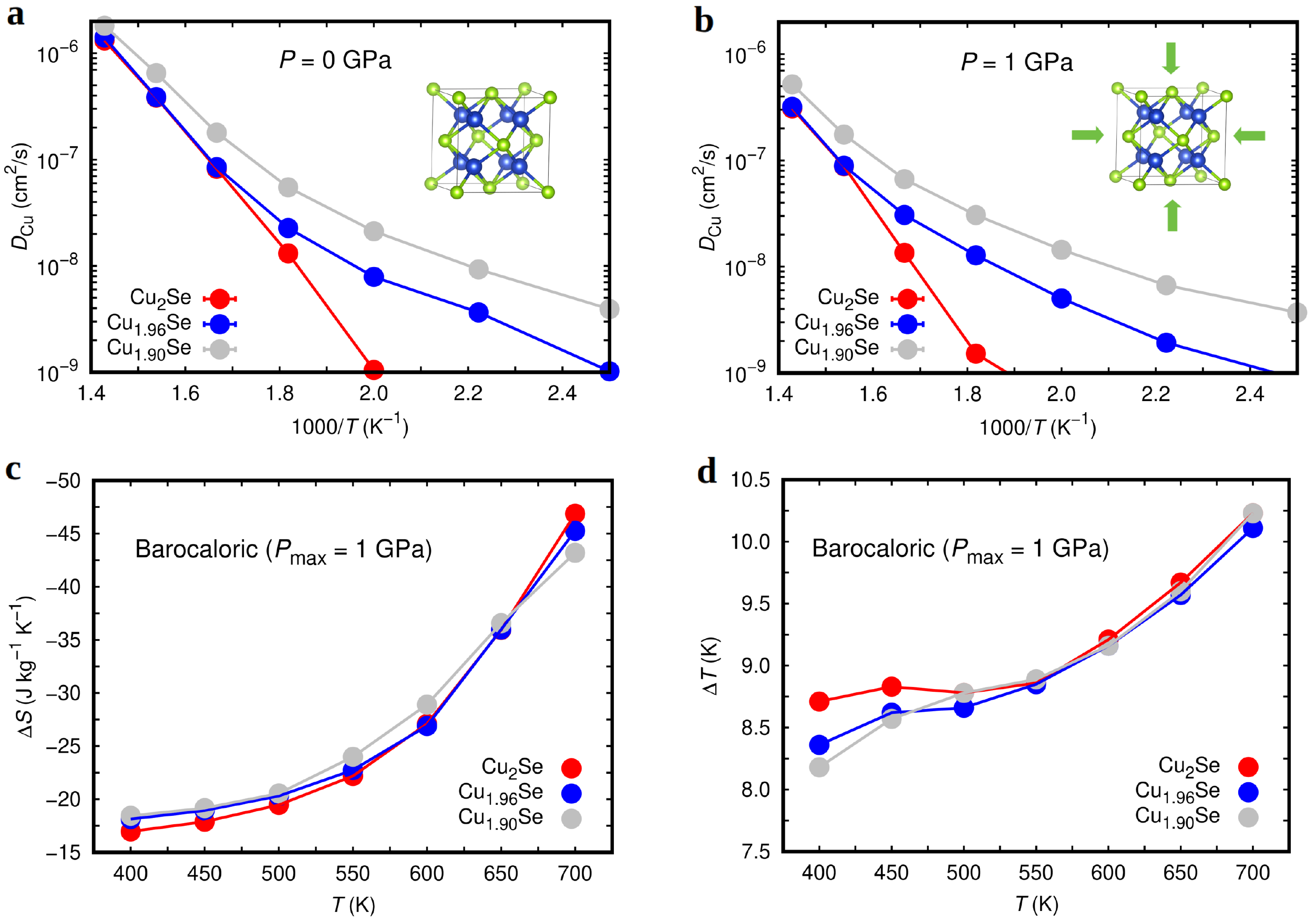}}
\caption{Barocaloric effects in bulk Cu$_{2-\delta}$Se at temperatures $400 \le T \le 700$~K.
	{\bf a} Ionic diffusion coefficients calculated at zero pressure and {\bf b} $P = 1$~GPa. 
        {\bf c} Isothermal entropy changes induced by a maximum hydrostatic pressure of $1$~GPa.  
	{\bf d} Adiabatic temperature changes induced by a maximum hydrostatic pressure of $1$~GPa.
	}
\label{fig2}
\end{figure*}

\subsection{Barocaloric effects}
\label{subsec:baro}
The normal ($\alpha$) to superionic ($\beta$) phase transition in Cu$_{2}$Se is very promising from a barocaloric 
point of view since it involves a large change of entropy (that is, a latent heat of $|\Delta S| = 78.4$~J~kg$^{-1}$~K$^{-1}$ 
\cite{murray74}) and it is sensitive to hydrostatic pressure, $P$. In particular, the critical 
$\alpha$~$\to$~$\beta$ transition temperature decreases under compression according to the numerical 
relationship $T_{\alpha \to \beta} = 408 - 0.624 P - 0.012 P^{2}$ (where temperature is in units of K 
and pressure of kbar) \cite{clark70}. A rough estimation of the potential barocaloric effects associated 
to such a first-order transition based on the experimentally reported values of the heat capacity ($C_{P} 
\approx 400$~J~kg$^{-1}$~K$^{-1}$ \cite{liu12}) and latent heat, that is, $|\Delta T| = -\frac{T}{C_{P}} 
|\Delta S|$ \cite{moya14}, leads to a colossal adiabatic temperature change of $\sim 80$~K. This value is 
much larger than the $|\Delta T|$ measured in the archetypal fast-ion conductor AgI with differential 
scanning calorimetry techniques ($36$~K), in which an analogous order-disorder phase transformation occurs 
close to room temperature \cite{cazorla17a}. Nevertheless, the exact nature of the low-$T$ phase in 
Cu$_{2}$Se has not been determined yet unequivocally and consequently we cannot simulate that phase with 
reliability. Hence, we are not going to analyse here the barocaloric effects associated with the 
$\alpha$~$\to$~$\beta$ phase transition. Experimental investigations on such potentially colossal 
barocaloric effects are in fact highly desirable. 

Meanwhile, large barocaloric effects have been recently predicted for the superionic conductor Li$_{3}$N
near room temperature, in which no structural phase transition occurs when applying moderate hydrostatic 
pressures on it \cite{cazorla18}. In particular, large isothermal entropy changes of about $25$~J~kg$^{-1}$~K$^{-1}$
have been estimated at $T = 300$~K and $P = 1$~GPa, which result from stress-induced variations on the 
volume an ion-transport properties of the material. Here, we investigate the possible existence of barocaloric 
effects in $\beta$--Cu$_{2}$Se at high temperatures, $400 \le T \le 700$~K, caused by similar atomistic 
mechanisms than in Li$_{3}$N. 

\begin{figure*}[t]
\centerline{
\includegraphics[width=1.00\linewidth]{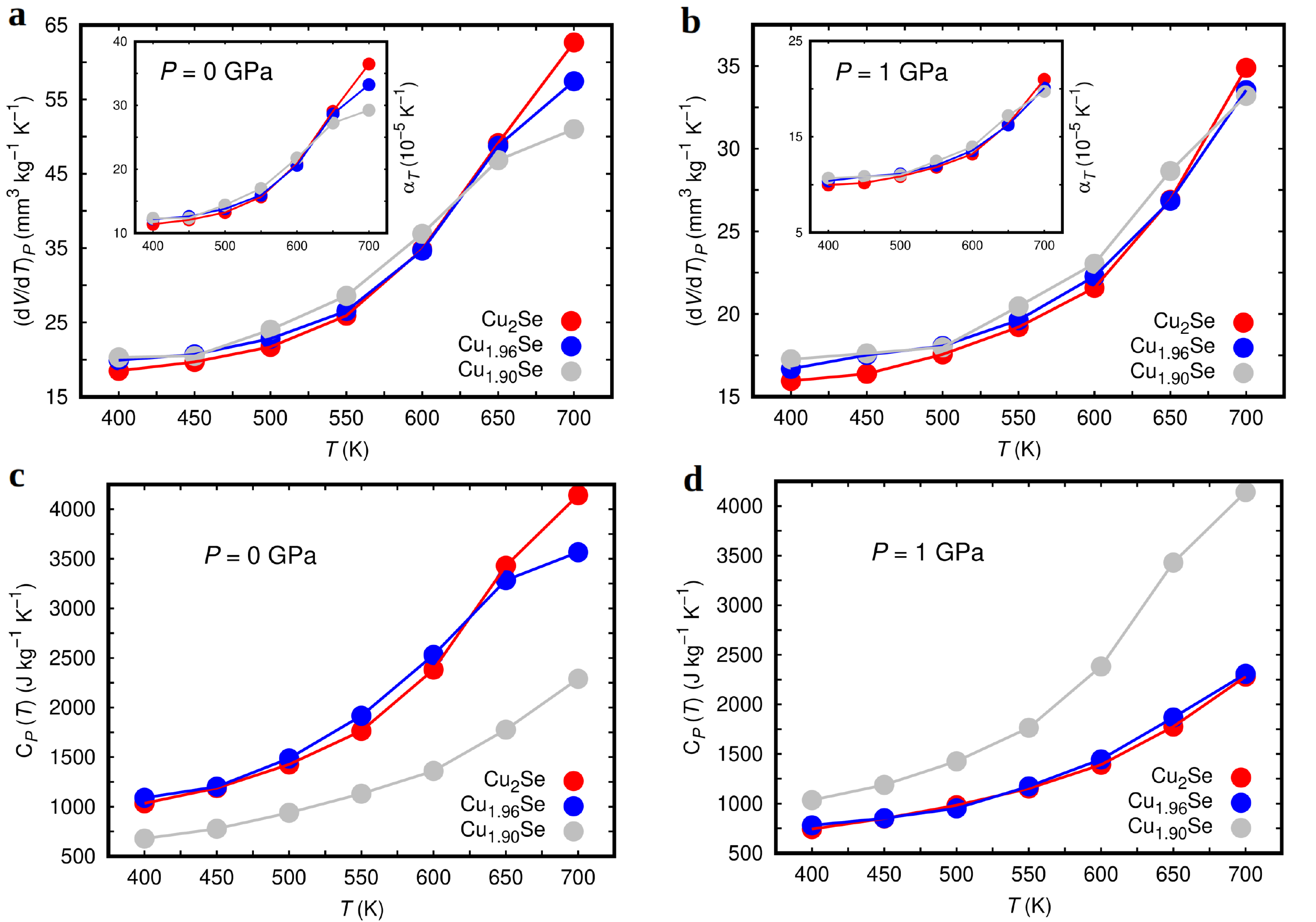}}
\caption{Thermodynamic properties of Cu$_{2-\delta}$Se at temperatures $400 \le T \le 700$~K.
         {\bf a} $T$--derivative of the volume calculated at zero pressure and {\bf b} $P = 1$~GPa.
	 Thermal expansion coefficients defined as $\alpha_{T} = \frac{1}{V} \left( \frac{dV}{dT} \right)_{P}$ 
	 are shown in the insets. {\bf c} Heat capacity calculated at zero pressure and {\bf d} $P = 
	 1$~GPa.
	 }
\label{fig3}
\end{figure*}

Figures~\ref{fig2}a-b show the influence of hydrostatic pressure on the diffusion coefficient of copper 
ions in $\beta$--Cu$_{2}$Se, for which we have considered several representative compositions \cite{clark70}. 
In all the cases, ionic transport is significantly depleted under an isotropic compression of $1$~GPa. 
For instance, at $T = 700$~K and zero pressure we estimate $1.3 \cdot 10^{-6}$ and $1.8 \cdot 10^{-6}$~cm$^{2}$s$^{-1}$ 
for Cu$_{2}$Se and Cu$_{1.90}$Se, respectively, whereas at the same temperature and $P = 1$~GPa we obtain 
$3.1 \cdot 10^{-7}$ and $5.2 \cdot 10^{-7}$~cm$^{2}$s$^{-1}$. Such large reductions in $D_{\rm Cu}$ suggest 
the presence of large entropy variations as induced by pressure; also, they indicate a potential enhancement 
in the overall thermodynamic stability of the material. Figure~\ref{fig2}c reports the isothermal entropy 
changes that we have explicitly calculated in $\beta$--Cu$_{2}$Se by using Eq.(\ref{eq:baro}) and considering 
a maximum pressure of $1$~GPa. Such estimated entropy changes in fact turn out to be quite large. Specifically, 
we obtain negative values of $|\Delta S| \sim 15$--$45$~J~kg$^{-1}$~K$^{-1}$ throughout the selected temperature 
interval, in which the largest entropy variations are attained at the highest considered $T$. 

Besides the large $D_{\rm Cu}$ variations caused by pressure, the intrinsically high anharmonicity of $\beta$--Cu$_{2}$Se 
\cite{danilkin12,voneshen17} appears to contribute also significantly to the estimated isothermal entropy changes
(here, we somewhat arbitrarily define ``anharmonicity'' as any other effect different from ionic diffusivity). 
By carrying out additional MD simulations in which non-zero ionic charges were adopted, and thus avoiding the 
appearance of ionic diffusion in the system (Sec.\ref{sec:results}), we found that the magnitude of the resulting 
barocaloric effects decreased by $\sim 60$\% of their original values. Therefore, we conclude that the influence 
of thermodynamic mechanisms different from Cu diffusion on the barocaloric performance of $\beta$--Cu$_{2}$Se is 
about $40$\%.    

Figure~\ref{fig2}d shows the adiabatic temperature changes estimated directly with Eq.(\ref{eq:deltat}) in 
the high-$T$ phase of copper selenide by considering a maximum compression of $1$~GPa. Large positive 
$\Delta T$ values of $8.50$--$10.25$~K are obtained within the selected temperature range. The calculated 
adiabatic temperature shifts increase steadily with temperature for $T \ge 500$~K and are slightly larger in 
the stoichiometric system at temperatures below that point. In Sec.\ref{sec:discussion}, we will discuss and 
compare the magnitude of the barocaloric effects predicted for $\beta$--Cu$_{2}$Se with those known from other 
materials reported in the literature.

In Fig.\ref{fig3}, we represent the value of the estimated $T$-induced volume variations and heat capacity as a 
function of temperature, pressure and composition. Such properties are used directly for computing $\Delta S$ 
and $\Delta T$ (Eqs.(\ref{eq:baro}) and (\ref{eq:deltat})), hence they are very relevant from a numerical point 
of view. The variation of the volume with respect to temperature behaves quite regularly as a function of $T$ and 
$P$. In particular, $\left(\frac{\partial V}{\partial T}\right)_{P}$ displays a parabolic-like temperature dependence 
and it decreases under pressure (Figs.\ref{fig3}a-b). The heat capacities computed for $\beta$--Cu$_{2}$Se and Cu$_{1.96}$Se 
exhibit similar dependences on $P$ and $T$ than their corresponding $T$-derivatives of the volume (Figs.\ref{fig3}c-d). 
However, the $C_{P}$ of $\beta$--Cu$_{1.90}$Se presents an anomalous behaviour since it increases noticeably under 
compression (we tentatively ascribe this tendency to some sort of effective interaction between point defects that is  
induced by $P$). We note that the results enclosed in Figs.\ref{fig3}c-d indicate a need for explicitly considering 
the $P$-dependence of $C_{P}$ in the calculation of $\Delta S$ and $\Delta T$, since this quantity may vary quite 
broadly under pressure (for instance, at $700$~K we calculate $2280$ and $4141$~J~kg$^{-1}$~K$^{-1}$ for the 
stoichiometric system at $1$~GPa and zero pressure, respectively). 

In the insets of Figs.\ref{fig3}a-b, we enclose the thermal expansion coefficients, $\alpha_{T} = \frac{1}{V} 
\left( \frac{dV}{dT} \right)_{P}$, calculated for $\beta$--Cu$_{2}$Se at different pressures, temperatures and 
compositions. In general, we obtain much larger $\alpha_{T}$ values than measured experimentally at zero pressure 
\cite{liu12}. For instance, at $T = 400$~K we compute $11.4 \cdot 10^{-5}$~K$^{-1}$, which is about $5$ times 
larger than the corresponding value determined in the experiments. A possible cause for these discrepancies may
be the neglection of electrostatic cohesion in our MD simulations. We note that such a $\alpha_{T}$ overestimation 
is likely to propagate into a certain overestimation of the $\Delta S$ values reported in this work. Regarding 
the heat capacity, our calculations also tend to amplify considerably this quantity. For instance, at $T = 600$~K 
and zero pressure we compute $2383$~J~kg$^{-1}$~K$^{-1}$, which is about $6$ times larger than the corresponding 
value determined in the experiments. 

By simultaneously considering the molecular dynamics overestimations of both $\alpha_{T}$ and $C_{P}$, however, 
we may conclude that our reported $\Delta T$ results should be pretty accurate due to a cancellation error between 
those two quantities (Eq.(\ref{eq:deltat})). Moreover, at high temperatures our $\Delta T$ results may be regarded 
as a lower bound of the adiabatic temperature changes that can be achieved in practice. For instance, at $T = 700$~K 
the simulated $\alpha_{T} / C_{P}$ ratio turns out to be three times smaller than the corresponding experimental 
value \cite{liu12}. Therefore, in spite of the inevitable shortcomings deriving from the use of classical potentials 
for simulating materials \cite{cazorlarev1,cazorlarev2}, we are confident that the $\Delta T$ results reported in 
this work are reliable.

\subsection{Elastocaloric effects}
\label{subsec:elasto}
\begin{figure*}[t]
\centerline{
\includegraphics[width=1.00\linewidth]{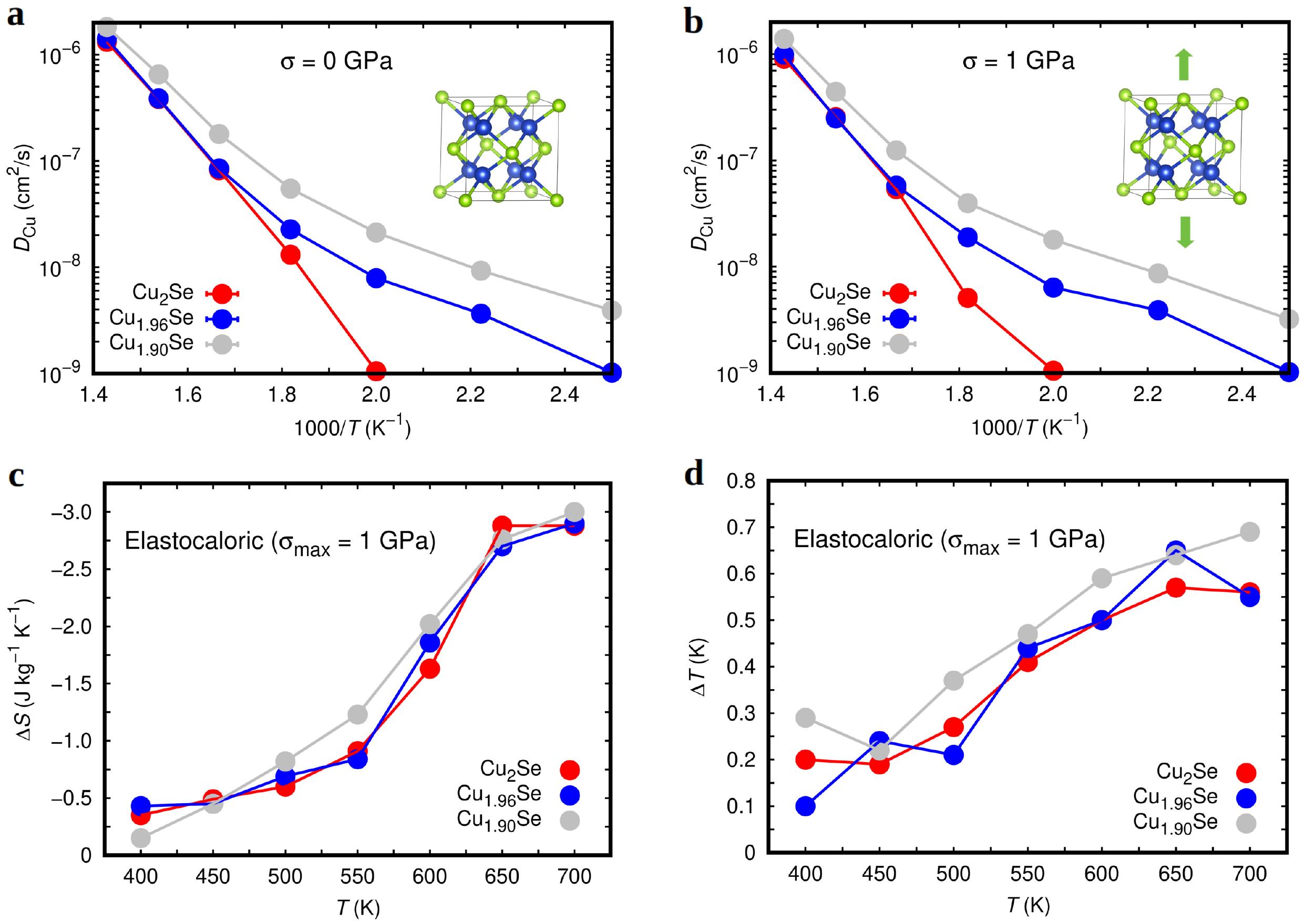}}
\caption{Elastocaloric effects in bulk Cu$_{2-\delta}$Se at temperatures $400 \le T \le 700$~K.
        {\bf a} Ionic diffusion coefficients calculated at zero uniaxial tensile stress and {\bf b} $\sigma 
	= 1$~GPa. {\bf c} Isothermal entropy changes induced by a maximum uniaxial tensile stress of 
	$1$~GPa. {\bf d} Adiabatic temperature changes induced by a maximum uniaxial tensile stress of 
	$1$~GPa.
         }
\label{fig4}
\end{figure*}

We have also investigated the potential of $\beta$--Cu$_{2}$Se as a candidate elastocaloric material. In particular, 
we have analysed the caloric response of the crystal under a maximum uniaxial compressive and tensile load of $1$~GPa
at different temperatures. Figure~\ref{fig4} shows the numerical $\Delta S$ and $\Delta T$ values obtained from  
molecular dynamics simulations in which we have considered uniaxial tensile stresses. The size of the estimated adiabatic 
temperature and isothermal entropy changes are about one order of magnitude smaller than calculated in the barocaloric 
case. For example, at $T = 400$~K we compute $\Delta S = -0.35$~J~kg$^{-1}$~K$^{-1}$ and $\Delta T = 0.20$~K for 
stoichiometric $\beta$--Cu$_{2}$Se (to be compared with $\Delta S = -17.5$~J~kg$^{-1}$~K$^{-1}$ and $\Delta T = 8.7$~K 
obtained in the barocaloric case). We note that the predicted elastocaloric effects present a practically negligible 
dependence on composition (Figs.\ref{fig4}b-d). Very similar $\Delta S$ and $\Delta T$ values have been obtained for 
the case of considering compressive uniaxial stresses in the simulations, which are not shown here. 

Figures~\ref{fig4}a-b show the influence of uniaxial tensile stress, $\sigma$, on the copper ionic diffusion coefficient 
of $\beta$--Cu$_{2}$Se at several temperatures. As it is appreciated therein, under a load of $1$~GPa the estimated
$D_{\rm Cu}$'s hardly change in comparison to the values obtained at zero stress. For example, for stoichiometric Cu$_{2}$Se 
at $T = 700$~K and $\sigma = 0$ we estimate $1.3 \cdot 10^{-6}$~cm$^{2}$s$^{-1}$ whereas at the same temperature and $\sigma = 
1$~GPa we obtain $1.0 \cdot 10^{-6}$~cm$^{2}$s$^{-1}$. Such a small $\sigma$-induced reduction in the ionic diffusivity 
explains the minuteness and sign of the elastocaloric effects predicted for $\beta$--Cu$_{2}$Se. Consequently, we may 
conclude that thermoelectric compounds presenting high ionic disorder in principle are not promising elastocaloric materials.

\begin{table*}
\centering
\begin{tabular}{c c c c c c c c}
\hline
\hline
$ $ & $ $ & $ $ & $ $ & $ $ & $ $ & $ $ & $ $ \\
	$ $ \quad & \quad  $T$ \quad & \quad $ P $ \quad & \quad $|\Delta S|$ \quad & \quad $|\Delta T|$ \quad & \quad $|\Delta T|/ P $ \quad & \quad ${\rm Materials~type}$ \quad & \quad $ {\rm Reference} $ \quad \\
$ $ \quad & \quad ${\rm (K)}$ \quad & \quad ${\rm (GPa)}$ \quad & \quad ${\rm (J K^{-1} Kg^{-1})}$ \quad & \quad ${\rm (K)}$ \quad & \quad ${\rm (K~GPa^{-1})} $ \quad & \quad $ $ \quad & \quad $ $ \quad\\
$ $ & $ $ & $ $ & $ $ & $ $ & $ $ & $ $ & $ $ \\
\hline
$ $ & $ $ & $ $ & $ $ & $ $ & $ $ & $ $ & $ $ \\
${\rm Ni_{51}Mn_{33}In_{16}}          $ \qquad & $ 330 $ & $ 0.25 $ & $ 41.0  $ & $ 4.0  $ & $ 16.0 $  & $ {\rm SMA} $ &  \cite{taulats15a}  \\
${\rm Fe_{49}Rh_{51}}                 $ \qquad & $ 310 $ & $ 0.11 $ & $ 12.5  $ & $ 8.1  $ & $ 73.6 $  & $ {\rm SMA} $ &  \cite{taulats15}  \\
${\rm BaTiO_{3}}                      $ \qquad & $ 400 $ & $ 0.10 $ & $ 2.4   $ & $ 1.0  $ & $ 10.0 $  & $ {\rm FE}  $ &  \cite{taulats16} \\
${\rm (NH_{4})_{2}SO_{4}}             $ \qquad & $ 220 $ & $ 0.10 $ & $ 130.0 $ & $ 8.0  $ & $ 80.0 $  & $ {\rm FE}  $ &  \cite{lloveras15} \\
${\rm [TPrA][Mn(dca)_{3}]}            $ \qquad & $ 330 $ & $ 0.01 $ & $ 30.5  $ & $ 4.1  $ & $ 410.0$  & $ {\rm OIH} $ & \cite{bermudez17} \\
${\rm [FeL_{2}][BF_{4}]_{2}}          $ \qquad & $ 262 $ & $ 0.03 $ & $ 80.0  $ & $ 3.0  $ & $ 100.0 $ & $ {\rm MC}  $ & \cite{vallone19} \\
${\rm (CH_{3})_{2}C(CH_{2}OH)_{2}}    $ \qquad & $ 320 $ & $ 0.52 $ & $ 510.0 $ & $ 45.0 $ & $ 86.5  $ & $ {\rm MC}  $ & \cite{li19,lloveras19} \\
${\rm AgI}                            $ \qquad & $ 400 $ & $ 0.25 $ & $ 62.0  $ & $ 36.0 $ & $ 144.0$  & $ {\rm FIC} $ & \cite{cazorla17a} \\
${\rm Li_{3}N}                        $ \qquad & $ 400 $ & $ 1.00 $ & $ 33.5  $ & $ 5.4  $ & $ 5.4  $  & $ {\rm FIC} $ & \cite{cazorla18} \\
${\rm Cu_{2}Se}                	      $ \qquad & $ 400 $ & $ 1.00 $ & $ 17.5  $ & $ 8.7  $ & $ 8.7  $  & $ {\rm TE/FIC} $ & $ {\rm This~work} $ \\
$ $ & $ $ & $ $ & $ $ & $ $ & $ $ & $ $ & $ $ \\ 
\hline
\hline
\end{tabular}
\label{tab:mcperform}
	\caption{Materials presenting large ($|\Delta T| > 1$~K) or giant ($|\Delta T| > 10$~K) barocaloric 
	effects at or near room temperature. $T$ represents working temperature, $P$ applied pressure, $|\Delta S|$ 
	isothermal entropy change, $|\Delta T|$ adiabatic temperature change, $|\Delta T|/P$ barocaloric strength, 
	``SMA'' shape-memory alloy, ``FE'' ferroelectric, ``OIH'' organic-inorganic hybrid perovskite, ``MC'' 
	molecular crystal, ``FIC'' fast-ion conductor, and ``TE'' thermoelectric.}
\end{table*}

\section{Discussion}
\label{sec:discussion}
To date, large barocaloric effects have been experimentally measured in a number of shape-memory alloys 
\cite{taulats15a,taulats15}, polar compounds \cite{taulats16,lloveras15}, organic-inorganic hybrid 
perovskites \cite{bermudez17}, molecular crystals \cite{vallone19,li19,lloveras19}, and the archetypal 
fast-ion conductor AgI \cite{cazorla17a}. In Table~I, we report several representative barocaloric compounds 
along with some of their basic cooling features measured at or near room temperature (made the exception 
of Li$_{3}$N \cite{cazorla18}, for which barocaloric experiments have not been performed yet). The 
thermoelectric fast-ion conductor $\beta$--Cu$_{2}$Se in fact turns out to be competitive with those 
archetypal materials in terms of barocaloric performance. 

The isothermal entropy change induced by a hydrostatic pressure of $1$~GPa in $\beta$--Cu$_{2}$Se at 
$T = 400$~K is relatively small as compared to those measured, for instance, in the shape-memory alloy
Ni$_{51}$Mn$_{33}$In$_{16}$, ferroelectric salt (NH$_{4}$)$_{2}$SO$_{4}$, molecular crystal 
(CH$_{3}$)$_{2}$C(CH$_{2}$OH)$_{2}$, and fast-ion conductor AgI (Table~I). On the other hand, the 
$|\Delta S|$ estimated for copper selenide is larger than or similar in magnitude to those found in 
the shape-memory alloy Fe$_{49}$Rh$_{51}$ and archetypal ferroelectric BaTiO$_{3}$. In terms of adiabatic
temperature shift, which arguably is the most important quality of caloric materials, $\beta$--Cu$_{2}$Se 
is better positioned. The calculated $|\Delta T|$ of $8.7$~K is larger than most values reported in 
Table~I, made the remarkable exception of the molecular crystal (CH$_{3}$)$_{2}$C(CH$_{2}$OH)$_{2}$ 
and fast-ion conductor AgI. It should be mentioned, however, that the experimental barocaloric effects 
that we review here have been obtained for hydrostatic pressures significantly smaller than $1$~GPa. 
For this reason, the predicted barocaloric strength of $\beta$--Cu$_{2}$Se, which is defined as the 
$|\Delta T|/P$ ratio, is not particularly exceptional ($\sim 10$~K GPa$^{-1}$).  

It is interesting to compare the barocaloric performance of $\beta$--Cu$_{2}$Se with that of Li$_{3}$N 
since both compounds are superionic and the corresponding $|\Delta S|$ and $|\Delta T|$ values have been 
obtained with similar computational methods \cite{cazorla18}. The isothermal entropy change estimated in 
Li$_{3}$N is practically two times larger than the value found in $\beta$--Cu$_{2}$Se under same $T$ and 
$P$ conditions. The main reason for such a difference is that the ionic diffusivity in Li$_{3}$N is much 
larger than that in $\beta$--Cu$_{2}$Se ($10^{-3}$ versus $10^{-9}$~cm$^{2}$s$^{-1}$, respectively), 
and thus the changes in ionic mobility caused by compression also result larger in Li$_{3}$N. Nevertheless, 
the adiabatic temperature change predicted for $\beta$--Cu$_{2}$Se is $\sim 60$\% larger than for Li$_{3}$N. 
This outcome is essentially due to the much smaller heat capacity estimated for the thermoelectric crystal 
(namely, the $C_{P}$ of Li$_{3}$N is about an order of magnitude larger than that of $\beta$--Cu$_{2}$Se 
\cite{cazorla18}). In view of these results, we propose that an effective design strategy for enhancing 
likely mechanocaloric effects in fast-ion conductors may consist in optimizing the corresponding heat 
capacities (that is, making them as small as possible by means of composition, for instance). 

It is worth reminding that the ``giant'' barocaloric effects (i.e., $|\Delta T| > 10$~K) measured in the also 
fast-ion conductor AgI (Table~I) were obtained for its normal to superionic phase transition, which is of 
first-order type and has associated a large latent heat \cite{cazorla17a}. Copper selenide also presents a 
$\alpha$~$\to$~$\beta$ phase transition near room temperature \cite{clark70} but for the reasons explained 
above (e.g., indetermination of the corresponding low-$T$ phase) we have restricted our analysis here to 
mechanocaloric effects occurring in the high-$T$ phase. Such mechanocaloric effects turn out to be large 
and are mostly due to continuous changes in the ionic diffusivity caused by pressure. Actually, the kind of 
second-order like changes disclosed here for $\beta$--Cu$_{2}$Se present great prospects in the context of 
refrigeration-cycle reversibility owing to the likely absence of mechanical hysteresis effects deriving from 
the nucleation of order-parameter domains \cite{cazorla18}. Nevertheless, our rough estimation of $|\Delta T| 
\sim 80$~K for the normal to superionic phase transition in Cu$_{2}$Se (Sec.\ref{subsec:baro}), and the previous 
success achieved in AgI for the same type of transformation (Table~I) \cite{cazorla17a}, should motivate also 
experimental searches of giant barocaloric effects in thermoelectric fast-ion conductor materials near 
room temperature.

\section{Conclusions}
\label{sec:conclusions}
Copper selenide and other similar silver- and sulfide-based compounds (e.g., Cu$_{2}$S, Cu$_{2}$Te, Ag$_{2}$Se, 
Ag$_{2}$S, Ag$_{2}$Te, and Cu$_{2-x}$Ag$_{x}$$X$ ($X$ = S, Se and Te) alloys) are thermoelectric fast-ion 
conductor materials for which huge thermoelectric figures of merit have been reported. However, owing to their 
ionic transport properties these materials are prone to suffer severe structural degradation when subjected to 
intense electric fields. In this computational work, we have shown that by applying moderate hydrostatic pressures 
of $\sim 1$~GPa it is possible to reduce significantly the ionic diffusivity of Cu$_{2}$Se, thus improving 
its thermodynamic stability, and to induce large barocaloric effects near room temperature. It is very likely 
that similar $P$-induced phenomena will occur also in analogous thermoelectric superionic materials. 

The caloric response of Cu$_{2}$Se is mostly originated by substantial changes on its ionic conductivity caused 
by compression. Such large barocaloric effects, namely, $|\Delta S| \sim 17.5$~J~kg$^{-1}$~K$^{-1}$ and $|\Delta T| 
= 8.7$~K at $T = 400$~K, are very promising for engineering novel solid-state cooling applications that do not 
require the application of electric fields. In this context, thermoelectric superionic materials are free 
of the degradation and energy-efficiency problems affecting solid-state refrigeration based on the Peltier effect. 
Moreover, we foresee the existence of giant barocaloric effects ($|\Delta T| \gg 10$~K) associated with the normal 
to superionic phase transition that occurs in Cu$_{2}$Se near room temperature (not simulated explicitly in this 
study). Hence, our theoretical findings on Cu$_{2}$Se should stimulate new energy-conversion experiments in 
thermoelectric fast-ion conductors that potentially can lead to robust and highly efficient solid-state cooling 
applications.

\section*{ACKNOWLEDGEMENTS}
Computational resources and technical assistance were provided by the Australian Government
and the Government of Western Australia through the National Computational Infrastructure
(NCI) and Magnus under the National Computational Merit Allocation Scheme and The Pawsey
Supercomputing Centre.

\end{document}